\documentclass[aps,pra,twocolumn,amsmath,amssymb,superscriptaddress,reprint,longbibliography]{revtex4-1}

\usepackage[utf8]{inputenc}
\usepackage{graphicx}
\usepackage{color}
\usepackage[colorlinks=true,citecolor=blue]{hyperref}
\usepackage{units}
\usepackage{amsmath}
\usepackage{tabularx}

\newcommand{\up}{\uparrow}
\newcommand{\down}{\downarrow}
\renewcommand{\vec}[1]{\mathbf{#1}}
\newcommand{\kB}{k_\text{B}}
\newcommand{\kBT}{k_\text{B}T}

\AtBeginDocument{%
    \newwrite\bibnotes
    \def\bibnotesext{Notes.bib}
    \immediate\openout\bibnotes=\jobname\bibnotesext
    \immediate\write\bibnotes{@CONTROL{REVTEX41Control}}
    \immediate\write\bibnotes{@CONTROL{%
    apsrev41Control,author="08",editor="1",pages="1",title="0",year="1"}}
     \if@filesw
     \immediate\write\@auxout{\string\citation{apsrev41Control}}%
    \fi
}%

\begin{document}
\title{Phase-dependent heat transport in Josephson junctions with p-wave superconductors and superfluids}
\author{Alexander G. Bauer}
\affiliation{Theoretische Physik, Universität Duisburg-Essen and CENIDE, D-47048 Duisburg, Germany}
\author{Björn Sothmann}
\affiliation{Theoretische Physik, Universität Duisburg-Essen and CENIDE, D-47048 Duisburg, Germany}
\date{\today}

\begin{abstract}
We investigate phase-coherent thermal transport in Josephson junctions made from unconventional superconductors or superfluids.
The thermal conductance is evaluated for one- and two-dimensional junctions within the Bogoliubov-de Gennes formalism.
We analyze three different scenarios of junctions between two triplet superconductors with (i) helical pairing, (ii) unitary chiral and (iii) nonunitary chiral pairing.
We find that the phase-dependent thermal conductance allows us to distinguish the different pairings and provides insight into the formation of topologically nontrivial Andreev bound states in the junction.
\end{abstract}

\maketitle
\section{\label{sec:intro}Introduction}
In 1957, BCS theory has been formulated as a microscopic theory of conventional superconductivity by Bardeen, Cooper and Schrieffer~\cite{bardeen_theory_1957}. It explains superconductivity in terms of the formation of spin-singlet $s$-wave Cooper pairs via an attractive, phonon-mediated electron-electron interaction. While it provides excellent agreement with experimental results for a large number of superconductors, it has become clear that there are other, unconventional superconductors which go beyond the original BCS framework and differ in the symmetry of their Cooper pair wave function~\cite{sigrist_phenomenological_1991}.

Unconventional superconductivity can be found, e.g., in high-temperature superconductors~\cite{bednorz_possible_1986} where Cooper pairs still form spin-singlet state but the orbital structure is of $d$-wave character~\cite{tsuei_pairing_2000}. Cooper pairs can also form spin-triplet states which are even under spin exchange and, thus, odd in the orbital degree of freedom due to $p$-wave pairing. Such triplet pairing occurs, e.g., in superfluid $^3$He~\cite{leggett_theoretical_1975,vollhardt_superfluid_2013}, Sr$_2$RuO$_4$~\cite{mackenzie_superconductivity_2003} and certain heavy-fermion superconductors~\cite{aoki_review_2019}.
In addition, spin-triplet pairing can also be created artifically in heterostructures where the proximity effect from a conventional BCS superconductor induces superconducting correlations in a nearby nonsuperconducting material~\cite{mcmillan_tunneling_1968}. The combination of spin-orbit interactions, exchange fields or external magnetic fields with broken translational invariance at interfaces then converts spin-singlet $s$-wave correlations into spin-triplet $p$-wave correlations~\cite{lutchyn_majorana_2010,oreg_helical_2010}. Similarly, $p$-wave superfluidity can also be induced in ultracold quantum gases~\cite{zhang_p_x+ip_y_2008}.

Recently, spin-triplet $p$-wave superconductors have received a lot of interest due to their nontrivial topological properties~\cite{ryu_topological_2010,sato_topological_2017}. In particular, they can host Majorana surfaces states~\cite{kitaev_unpaired_2001,kwon_fractional_2004,mourik_signatures_2012,he_chiral_2017,alicea_new_2012,beenakker_search_2013,aguado_majorana_2017} which have non-Abelian braiding statistics~\cite{alicea_non-abelian_2011} and are potential candidates for topologically protected qubits~\cite{nayak_non-abelian_2008}.
Furthermore, the nontrivial momentum dependence of $p$-wave order parameters allows for the realization of different types of triplet superconductivity such as helical pairing where the order parameter is of the form $\boldsymbol\sigma\cdot\vec p$, unitary chiral pairing with an order parameter of the form $p_x\pm ip_y$ and nonunitary chiral pairing with broken time-reversal symmetry where only one spin species is paired.

\begin{figure}[h!!]
	\includegraphics[width=0.9\columnwidth]{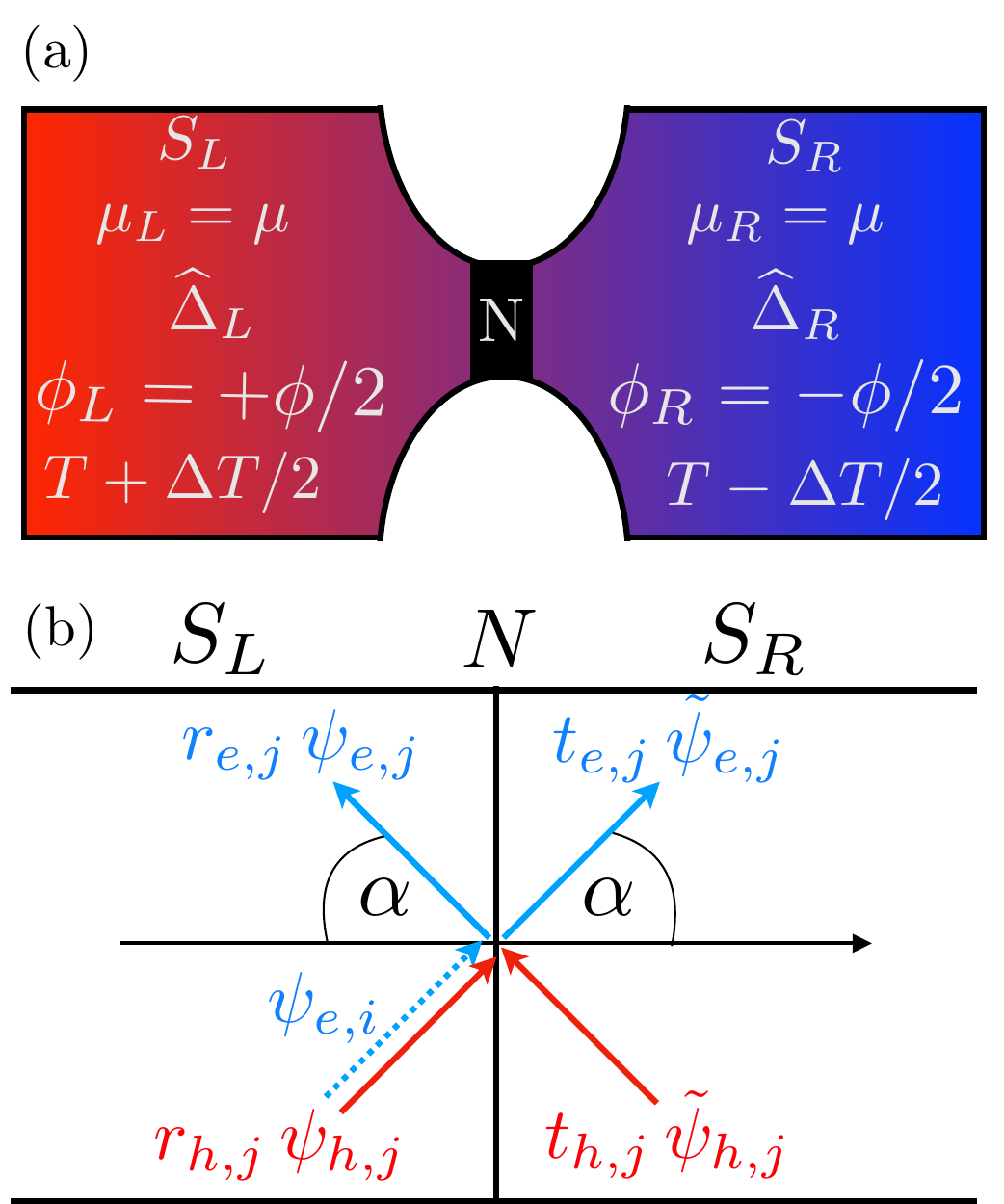}
	\caption{\label{fig:setup} (a) Short Josephson junction consisting of two superconducting reservoirs with order parameters $\hat{\Delta}_L$ and $\hat{\Delta}_R$ connected  by a normal region $N$. The two superconductors are at the same chemical potential $\mu$ while a temperature bias $\Delta T$ and phase bias $\phi$ are applied across the junction. (b) Scattering states: an incoming electronlike state $\psi_{e,i}$ (dashed blue line) impinging on the junction under an angle $\alpha$ gives rise to transmitted electronlike quasiparticles $t_{e,j}\tilde\psi_{e,j}$ (blue solid line) and holelike quasiparticles $t_{h,j}\tilde\psi_{h,j}$ (red solid line) as well as to reflected  electronlike quasiparticles $r_{e,j} \psi_{e,j}$ (blue solid line) and holelike quasiparticles $r_{h,j}\psi_{h,j}$ (red solid line). Note that the group velocity $\mathbf{v_g} \propto \partial_\mathbf{k} \varepsilon$ is parallel to the momentum vector for electronlike and antiparallel for holelike modes.} 
\end{figure}

It is, thus, desirable to probe, control and manipulate the properties of spin-triplet superconductors. This can be achieved, e.g., in SNS Josephson junctions where two superconductors are coupled to each other by a normal region. Andreev reflection at the superconductor-normal metal interfaces gives rise to the formation of Andreev bound states at discrete energies within the superconducting gap. In a junction of spin-triplet superconductors, topologically protected zero-energy Andreev bound states form which are Majorana bound states~\cite{fu_superconducting_2008,fu_josephson_2009}. They give rise to a $4\pi$-periodic Josephson effect that can be detected experimentally~\cite{rokhinson_fractional_2012,wiedenmann_4pi-periodic_2016,bocquillon_gapless_2017,deacon_josephson_2017,li_4pi-periodic_2018,laroche_observation_2019}.  This is, however challenging due to issues with quasiparticle poisoning~\cite{higginbotham_parity_2015,van_woerkom_one_2015} and the contribution of additional trivial modes.

Here, we propose phase-coherent heat transport in a temperature-biased Josephson junction as an alternative way of probing topologically nontrivial Josephson junctions made from spin-triplet $p$-wave superconductors. A phase-dependent contribution to the thermal conductance of a Josephson junction was predicted by Maki and Griffin~\cite{maki_entropy_1965,maki_entropy_1966} and subsequently been discussed for tunnel junctions~\cite{guttman_phase-dependent_1997,guttman_interference_1998} and superconducting point contacts~\cite{zhao_phase_2003,zhao_heat_2004}. Only very recently has the phase-dependent thermal conductance been measured experimentally~\cite{giazotto_phase-controlled_2012,giazotto_josephson_2012}.
It arises from Andreev-like processes where an electronlike (holelike) quasiparticle above the superconducting gap is transmitted or reflected as a holelike (electronlike) quasiparticle~\cite{kamp_phase-dependent_2019}. Thus, the phase-dependence can provide nontrivial information about Andreev bound states~\cite{sothmann_fingerprint_2016}. At the same time, phase-coherent thermal transport can also be useful for future caloritronic applications as it allows for the operation of heat interferometers~\cite{giazotto_phase-controlled_2012, giazotto_josephson_2012,martinez-perez_fully_2013,fornieri_nanoscale_2016,fornieri_0-pi_2017}  , thermal switches~\cite{sothmann_high-efficiency_2017}, thermal rectifiers and diodes~\cite{martinez-perez_efficient_2013,giazotto_thermal_2013,fornieri_normal_2014,martinez-perez_rectification_2015}, thermal transistors~\cite{giazotto_proposal_2014,fornieri_negative_2016}, heat circulators~\cite{hwang_phase-coherent_2018} and superconducting refrigerators~\cite{solinas_microwave_2016,hofer_autonomous_2016,vischi_thermodynamic_2019}.  

In this paper, we demonstrate that the thermal conductance of a Josephson junction can be used to distinguish spin-singlet and spin-triplet superconductors and elaborate the characteristic transport signatures of the triplet case. Furthermore, we elaborate in how far the thermal conductance can provide further insight into the nature of the spin-triplet state.  Compared to Andreev bound state spectroscopy~\cite{pillet_andreev_2010} which can also provide information about the superconducting pairing, thermal conductance measurements offer the advantage that they do not rely on three-terminal measurements and can, thus, be implemented more easily with superfluids or systems based on exotic superconducting materials. Furthermore, thermal conductance measurements rely on quasiparticle transport above the superconducting gap rather than Cooper pair transfer and, thus, provide different means of obtaining information about Andreev bound states compared to spectroscopy and Josephson currents.
We remark that while in the rest of the paper we will always use the term superconductor, our results apply to superfluids as well.

The paper is organized as follows. In Sec.~\ref{sec:model} we present our model of a spin-triplet $p$-wave Josephson junction and introduce the Bogoliubov-de Gennes formalism used for its theoretical description. We then discuss the phase-dependent thermal conductance of three types of junctions involving superconductors with helical pairing in Sec.~\ref{sec:helsc}, unitary chiral pairing in Sec.~\ref{sec:chisc}, and nonunitary chiral pairing in Sec.~\ref{sec:favchisc}. Conclusions are drawn in Sec.~\ref{sec:con}.

\section{\label{sec:model}Model}
We consider a two-dimensional Josephson junction built from two spin-triplet $p$-wave superconductors which are connected by a normal region, see Fig.~\ref{fig:setup}. 
Both superconductors are kept at the same chemical potential $\mu$. 
The order parameters $\hat{\Delta}_L$ and $\hat{\Delta}_R$ can in general be different and, thus, allow for a different orientation of the Cooper pair spin in the two superconductors.
The junction is biased by a temperature difference $\Delta T$ and a difference in the phases of the superconducting order parameters $\phi$. 
We describe the Josephson junction within the Bogoliubov-de Gennes (BdG) scattering formalism \cite{blonder_transition_1982}. In the basis
\begin{equation}
\left(u_\uparrow(\vec k), u_\downarrow(\vec k), v_\uparrow(\vec k), v_\downarrow(\vec k) \right)^T,
\end{equation}
where $u_{\sigma}$ and $v_{\sigma}$ represent electron- and holelike quasiparticle excitations with spin $\sigma$,
the BdG Hamiltonian in momentum space is given by \cite{gennes_superconductivity_1999}
\begin{equation}
H_{\text{BdG}}=
\begin{pmatrix}
\hat{H}_0(\vec k) & \hat{\Delta}(\vec k) \\
-\hat{\Delta}^\star(-\vec k) & -\hat{H}_0^\star(-\vec k) 
\end{pmatrix}.
\label{eq:hbdg}
\end{equation}
$H_0$ represents the quadratic single-particle Hamiltonian  which is isotropic in spin space and therefore given by
\begin{equation}
\hat{H}_0(\vec k)= \sigma_0 \otimes \left[ \frac{\hbar^2 \vec k^2}{2m} -\mu \right],
\label{eq:hSP}
\end{equation}
where $\sigma_0$ denotes the unit matrix in the spin space.
In contrast to conventional spin-singlet $s$-wave pairing, the order parameter of the triplet pairing $\hat{\Delta}(\vec k)$ has a matrix structure which can be parametrized by the Balian-Werthammer vector~\cite{balian_superconductivity_1963} $\vec d(\vec k)=(d_x(\vec k),d_y(\vec k),d_z(\vec k))^T$ and the vector of Paulimatrices $\boldsymbol{\sigma}=(\sigma_x,\sigma_y,\sigma_z)^T$ as
\begin{equation}
\hat{\Delta}(\vec k)=\begin{pmatrix}
\Delta_{\uparrow,\uparrow}(\vec k) & \Delta_{\uparrow,\downarrow}(\vec k) \\
 \Delta_{\downarrow,\uparrow}(\vec k) &  \Delta_{\downarrow,\downarrow}(\vec k)
\end{pmatrix}=\vec d(\vec k)\cdot\boldsymbol{\sigma}i\sigma_y.
\label{eq:PP}
\end{equation}
The  components $d_x(\vec k)$, $d_y(\vec k)$ and $d_z(\vec k)$ correspond to Cooper pairs with an $S_x=0,S_y=0$ and $S_z=0$ spin projection, respectively. In terms of the $d$-vector the pairing potential has the explicit form
\begin{equation}
\hat{\Delta}(\vec k)=\begin{pmatrix}
-d_x(\vec k)+id_y(\vec k) & d_z(\vec k) \\
d_z(\vec k) &  d_x(\vec k)+id_y(\vec k)
\end{pmatrix}.
\label{eq:Pexp}
\end{equation}
As the triplet correlations are even under spin exchange, the Pauli principle requires that the full wavefunction is either odd under the exchange of spatial coordinates or time arguments. Here, we focus on spin-triplet $p$-wave pairing which is even in the time domain. Due to this $p$-wave character, all triplet pairing correlations are odd under momentum inversion,
\begin{equation}
\vec d(\vec k)=-\vec d(-\vec k).
\label{eq:dp}
\end{equation}
We remark that alternatively the possibility of odd-frequency superconductivity arises where spin-triplet pairing occurs in an $s$-wave state~\cite{linder_odd-frequency_2017}.

Furthermore, the $\vec d$-vector is directly proportional to $\vec d(\vec k) \propto \Delta_0/k_F$, where $\Delta_0$ determines the size of the gap in the bulk energy spectrum at $|\vec k|=k_F$.

In our analysis we focus on short SNS junctions where the length of the normal region $L$ is short compared to the superconducting coherence length $\xi \sim \hbar v_F/|\Delta_0|$. In this case, it is possible to model the N region by a delta-shaped barrier at the interface between the two superconductors. Furthermore, we assume that the Fermi wavelength of the superconductor is much smaller then the one of the normal region $\lambda_{F,S} \ll \lambda_{F,N}$.
As a result, the order parameters of the superconducting regions quickly approach their bulk value at the SN interface and we may approximate their spatial dependence by a stepfunction behavior~\cite{likharev_superconducting_1979,beenakker_universal_1991,beenakker_specular_2006},
\begin{equation}
\hat{\Delta}(x)=\hat{\Delta}_L\Theta(-x)+\hat{\Delta}_R\Theta(+x).
\end{equation}
For a two-dimensional SNS junction, we define the angle of incidence $\alpha$ at the normal interface in terms of the momenta of left-moving and right-moving quasiparticles as
\begin{align}
	\vec k_r&=k_F(\cos\alpha,\sin\alpha),\\
	\vec k_l&=k_F(-\cos\alpha,\sin\alpha),
\end{align}
cf. also Fig.~\ref{fig:setup}(b). We remark that for the special case of a one-dimensional  junction $\alpha=0$.

We solve the BdG equation 
\begin{equation}
	H_\text{BdG}\psi_{e/h,i}e^{i\vec k_{l/r}\cdot \vec r}=\varepsilon_{\vec k}\psi_{e/h,i}e^{i\vec k_{l/r}\cdot \vec r},
	\label{eq:scstates}
\end{equation}
for a given energy above the gap $\varepsilon>|\Delta_0|$  in the Andreev regime $\mu \gg \varepsilon,|\Delta_0|$ to determine the scattering states $\psi_{e/h,i}e^{i\vec k_{l/r}\cdot\vec r}$ of the superconducting bulk. From Eq.~\eqref{eq:scstates} we obtain eight scattering states which are characterized as electronlike or holelike, left-moving or right-moving and by their spin index $i$.

For the left side of the junction ($x<0$) an incident right-moving electronlike mode $\psi_{e,i}e^{i\vec k_r\cdot\vec r}$ gives rise to a backscattered electronlike mode $\psi_{e,j}e^{i\vec k_l\cdot\vec r}$ or to an Andreev reflected holelike model $\psi_{h,j}e^{i\vec k_r\cdot\vec r}$,
\begin{equation}
	\Psi_L=\psi_{e,i}e^{i\vec k_r\cdot \vec r}+\sum_j \Big[r_{e,j}\psi_{e,j}e^{i\vec k_l\cdot \vec r}+r_{h,j}\psi_{h,j}e^{i\vec k_r\cdot\vec r}\Big].
\end{equation}
In addition, the incoming mode can also be transmitted to the right side ($x>0$) as an electronlike mode $\tilde\psi_{e,j}e^{i\vec k_r\cdot\vec r}$ or as a holelike mode $\tilde\psi_{h,j}e^{i\vec k_l\cdot\vec r}$,
\begin{equation}
	\Psi_R=\sum_j\Big[t_{e,j}\tilde\psi_{e,j}e^{i\vec k_r\cdot\vec r}+t_{h,j}\tilde\psi_{h,j}e^{i\vec k_l\cdot\vec r}\Big].
\end{equation}
We remark that the explicit shape of the electron- and holelike states on the left and right side can in general differ due to different order parameters $(\hat\Delta_L \neq \hat\Delta_R)$.
At the normal interface $(x=0)$ the boundary conditions are then given by
\begin{align}
\Psi_L&=\Psi_R,\\
\partial_x \Psi_R-\partial_x\Psi_L&=2Zk_F\Psi_L,
\label{eq:BC2}
\end{align}
where $Z$ is a dimensionless parameter related to the transmission in the normal state as
\begin{align}
\tau=\frac{1}{1+Z^2}.
\label{eq:tbarrier}
\end{align}
From the matching procedure we obtain the transmission function $\mathcal{T}=\sum_{j}\left[|t_{e,j}|^2+|t_{h,j}|^2 \right]$ through the junction. 

When a temperature bias $\Delta T$ is applied across the junction, the thermal conductance $\kappa$ in linear response is given by~\cite{butcher_thermal_1990} 
\begin{equation}
\kappa(\phi)=\frac{g}{h} \int_{|\Delta_0|}^{+ \infty} d\varepsilon\, \varepsilon\mathcal{T}(\varepsilon,\phi) \frac{df}{dT}.
\label{eq:tc}
\end{equation}
Here, $f=\left[\exp(\varepsilon/(\kBT)) +1\right]^{-1}$ denotes the Fermi distribution function  in equilibrium. The degeneracy factor $g$ accounts for spin degeneracy as well as particle-hole degeneracy. Hence, for the case of helical and unitary chiral pairing we have $g=4$ while in the nonunitary chiral case $g=1$.

\section{\label{sec:res}Results}
In the following, we are going to discuss the phase-dependent thermal conductance of Josephson junctions involving spin-triplet $p$-wave superconductors. In Sec.~\ref{sec:helsc}, we start our analysis with the simplest case of unitary helical pairing which is isotropic in $\vec k$ space. We then turn to the case of unitary but anisotropic chiral pairing where the relative orientation of the order parameters in the two superconductors becomes important in Sec.~\ref{sec:chisc}. Finally, we will investigate a junction with nonunitary chiral pairing where only one spin species is paired in Sec.~\ref{sec:favchisc}. While the main focus of our analysis lies on one-dimensional transport through the junction, we  also discuss qualitative differences arising in the two-dimensional case for all three pairing scenarios.

\subsection{\label{sec:helsc}Helical Pairing}
We first consider superconductors with helical pairing which preserves time-reversal symmetry. This type of pairing occurs intrinsically in the superfluid He$^3$ B-phase \cite{vollhardt_superfluid_2013}. Alternatively, it can be generated artifically in superconductor-topological insulator hybrid structures~\cite{fu_superconducting_2008,tkachov_suppression_2013}. These hybrid structures differ from our model in that they exhibit a linear dispersion which in turn gives rise to the effect of superconducting Klein tunneling~\cite{tkachov_helical_2013}.
The order parameter for helical pairing is characterized by a $d$-vector proportional to the momentum vector $\vec k$~\cite{balian_superconductivity_1963},
\begin{equation}
\vec d(\vec k)=\frac{\Delta_0}{k_F}\left(k_x,k_y,k_z\right)^T.
\label{eq:dhel}
\end{equation}
For an isolated bulk superconductor where the order parameter $\Delta_0$ is position independent and can be gauged real, the Hamiltonian~\eqref{eq:hbdg} preserves time-reversal symmetry
\begin{equation}
TH_{\text{BdG}}(\vec k)T^{-1}= H_{\text{BdG}}(-\vec k).
\end{equation}
In combination with the particle-hole symmetry of the BdG Hamiltonian, superconductors with helical pairing belong to the Cartan class DIII~\cite{ryu_topological_2010}. It is furthermore straightforward to demonstrate that the order parameter is nodeless and isotropic in $\vec k$-space,
\begin{align}
\frac{1}{2}\text{Tr}\left[\hat{\Delta}(\vec k)\hat{\Delta}^\dagger(\vec k)\right]=|\Delta_0|^2,
\end{align}

\begin{figure}
	\includegraphics[width=1.0\columnwidth]{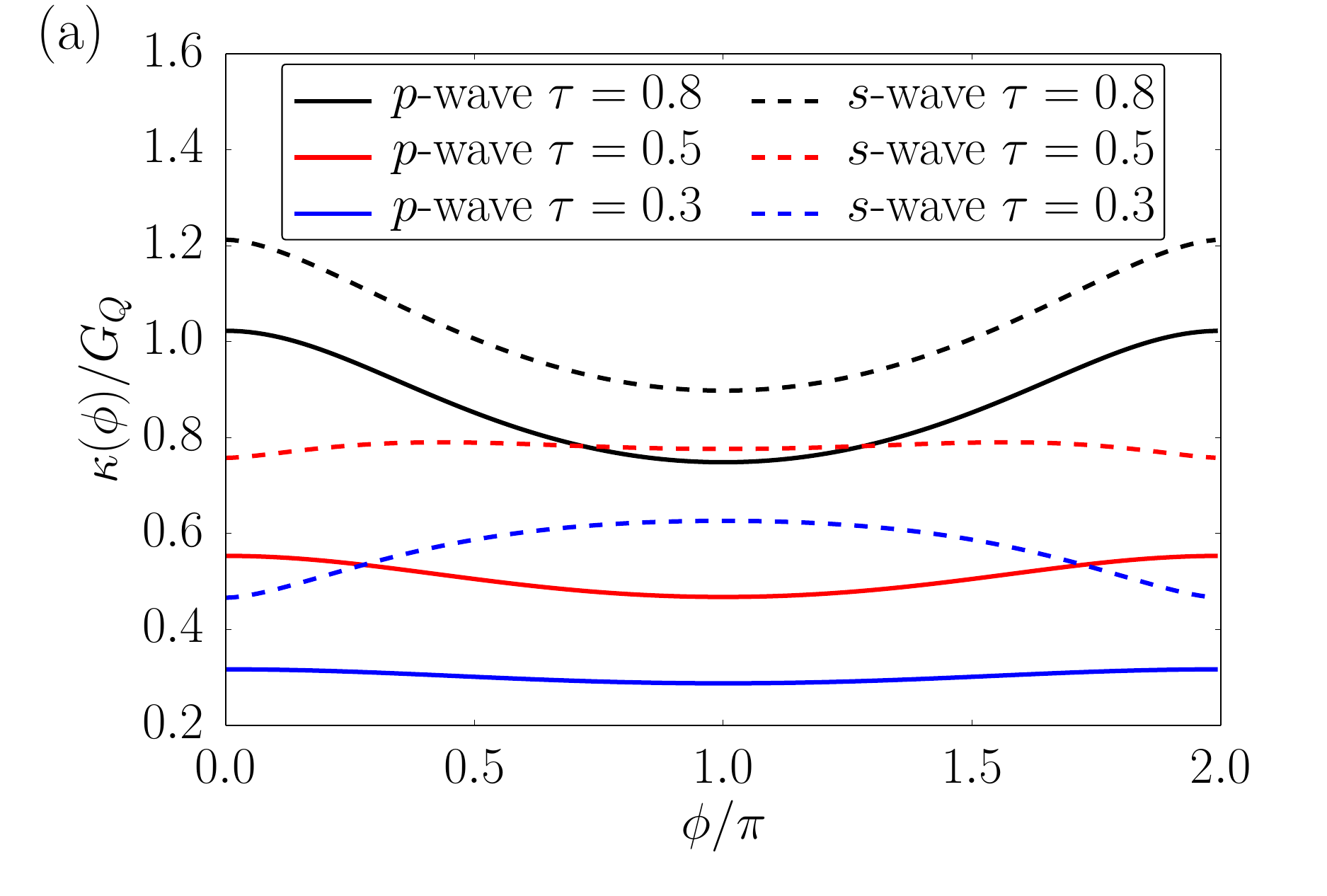}
	\includegraphics[width=1.0\columnwidth]{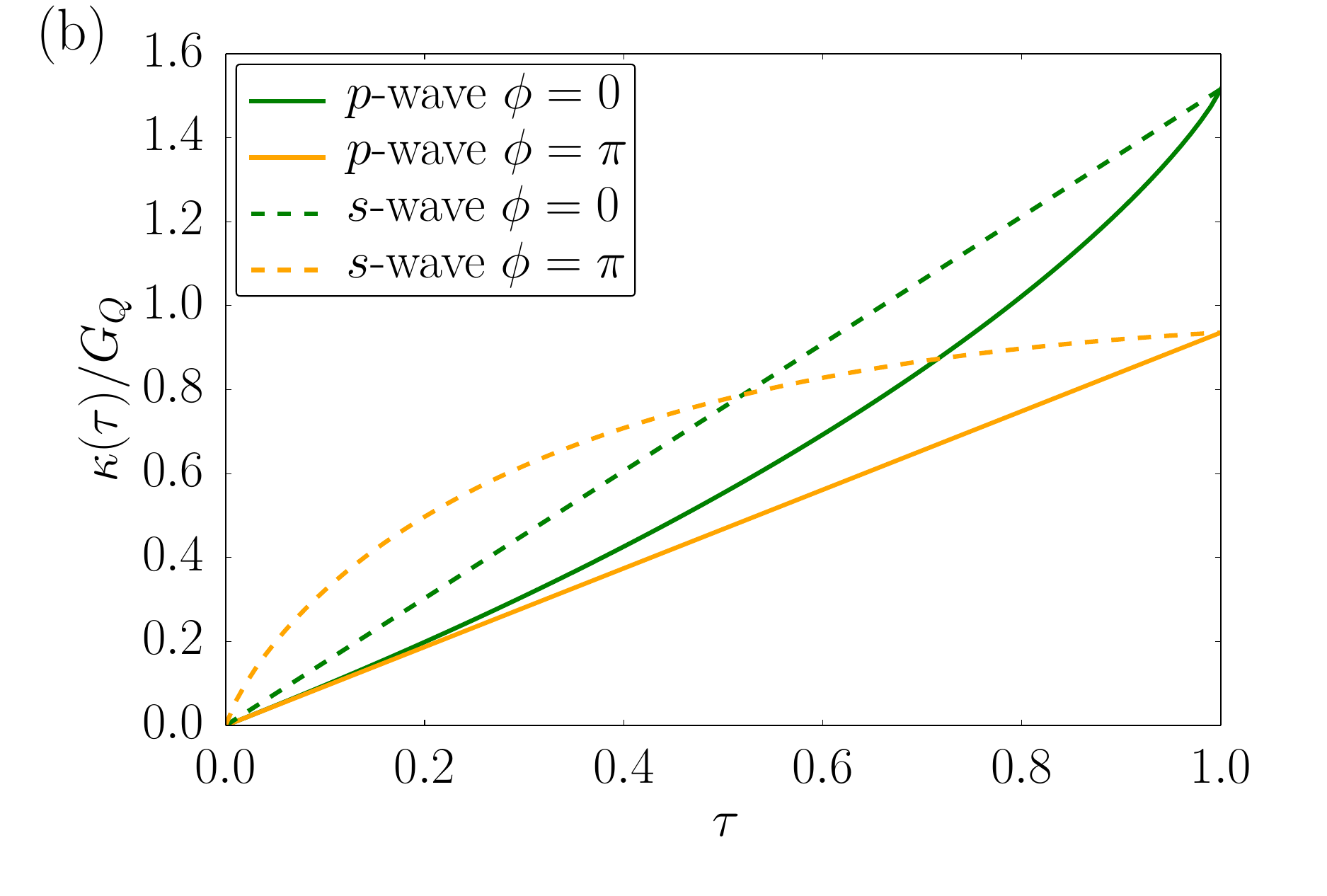}
	\caption{\label{fig:1dbhas} Thermal conductance $\kappa$ in units of the heat flux quantum $G_Q=\pi^2k_B^2T/(3h)$ for a one-dimensional SNS junction with helical pairing.  (a) Comparison of $\kappa(\phi)$ between the s- and helical $p$-wave case for varying transmission $\tau$. (b) Comparison of $\kappa(\tau)$ between the s- and $p$-wave case for $\phi=0$ and $\phi=\pi$. In contrast to the trivial $s$-wave pairing there is always a minimum of $\kappa$ at $\phi=\pi$ for the $p$-wave case independently of $\tau$. The temperature is set to $\kBT=|\Delta_0|/2$.}
\end{figure}
We start our analysis by considering the simple case of a one-dimensional SNS junction. Since the order parameter is isotropic, transport properties of the junction do not depend on its orientation. In the following, we thus consider transport along the $x$ axis such that the order parameter becomes diagonal in spin space and has $\up\up$ and $\down\down$ components only, cf. Eq.~\eqref{eq:PP},
\begin{equation}
\hat{\Delta}(\vec k)= \frac{\Delta_0}{k_F}
\left(\begin{array}{cc}
-k_x & 0 \\
0 & k_x
\end{array}\right).
\label{eq:dhel1d}
\end{equation}
We remark that Cooper pairs have no net spin because both $\up\up$ and $\down\down$ pairing are equally likely. Within the BdG formalism, we obtain the transmission function of the junctions,
\begin{equation}
\mathcal{T}(\varepsilon,\phi)= \frac{\varepsilon^2-|\Delta_0|^2}{\varepsilon^2/\tau-|\Delta_0|^2 \cos^2(\phi/2)},
\label{eq:TFH}
\end{equation}
where $\tau$ denotes the transmission in the normal state and $\phi$ is the phase difference between the two superconductors.

The resulting phase-dependent thermal conductance $\kappa$ is shown by solid lines in Fig.~\ref{fig:1dbhas}(a) for different values of $\tau$. For comparison, the corresponding thermal conductance of a conventional spin-singlet $s$-wave junction is indicated by dashed lines. We find that in both cases the thermal conductance is $2\pi$-periodic and decreases when the transmission $\tau$ is lowered. In the case of $p$-wave pairing, the thermal conductance is always minimal at $\phi=\pi$ independent of the value of $\tau$. This is in contrast to the case of $s$-wave pairing where a minimum at $\phi=\pi$ occurs only for large $\tau$ while in the tunneling limit $\tau\ll1$ the thermal conductance becomes maximal at $\phi=\pi$.

This difference becomes even clearer when plotting the thermal conductance at phase differences $\phi=0$ and $\phi=\pi$ as a function of the normal state transmission $\tau$, cf. Fig.~\ref{fig:1dbhas}(b). While in the case of $p$-wave pairing we find $\kappa(\phi=0)>\kappa(\phi=\pi)$ for all values of $\tau$, the two curves cross each other in the $s$-wave case for intermediate transmissions $\tau\approx 0.5$.

\begin{figure}
	\includegraphics[width=0.85\columnwidth]{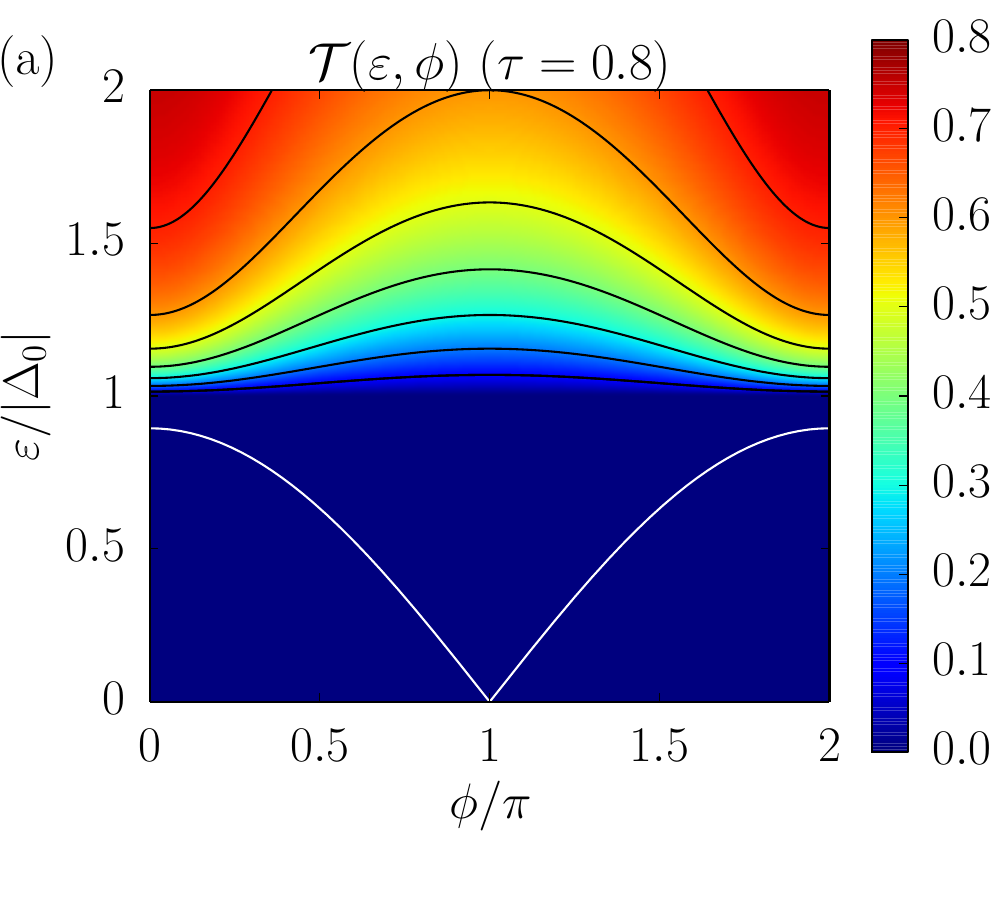}
	\includegraphics[width=0.85\columnwidth]{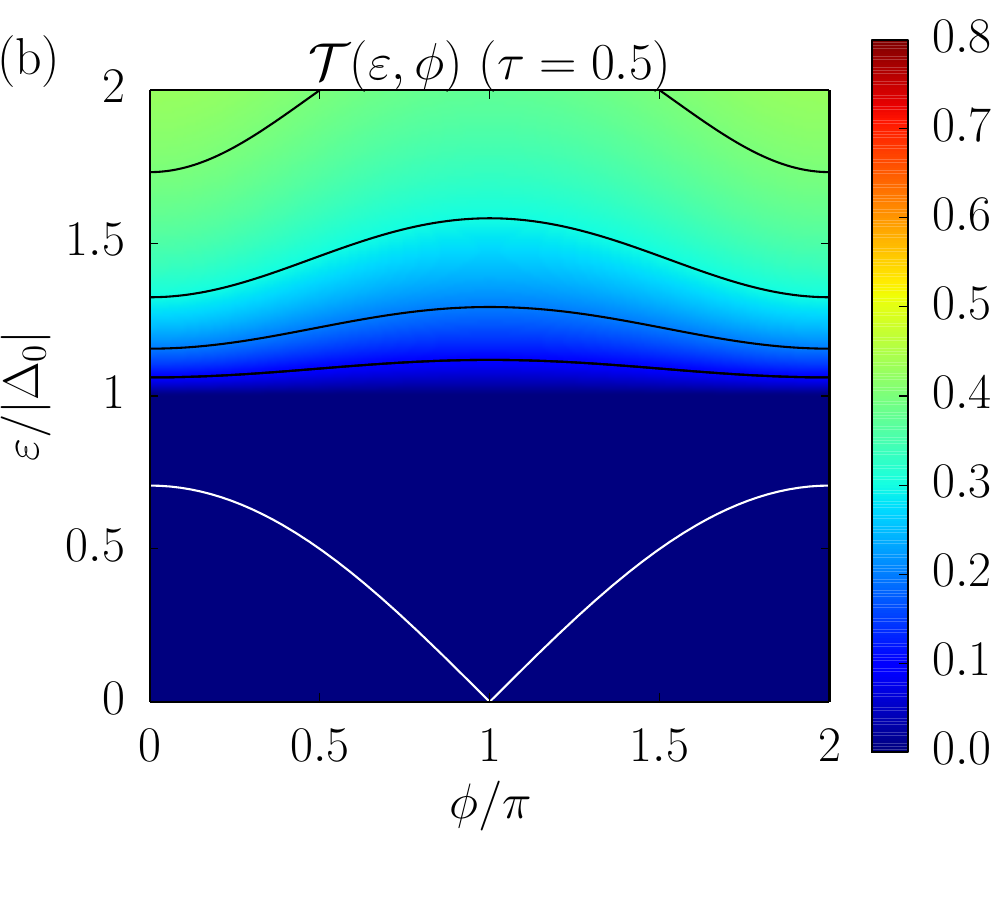}
	\caption{\label{fig:1dbhasDOS} 
    Transmission function $\mathcal{T}(\varepsilon,\phi)$ of a one-dimensional SNS junction with helical pairing for different transmissions  $\tau$. The presence of a topologically protected Majorana bound state at $\phi=\pi$ is connected to a minimum of the transmission above the superconducting gap.}
\end{figure}
To understand the underlying physics, let us now focus on the Andreev bound states forming inside the superconducting gap. In the case of a junction with spin-singlet $s$-wave superconductors, their energy is given by
\begin{equation}
	E^\text{ABS}_s(\phi)=\pm|\Delta_0|\sqrt{1-\tau\sin^2\frac{\phi}{2}}.
	\label{eq:ABSBswav}
\end{equation}
For transparent junctions, the Andreev bound state energy goes to zero as $\phi=\pi$. In consequence, spectral weight is removed from above the superconducting gap and the thermal conductance gets reduced. In contrast, for the tunneling limit the Andreev bound states at $\phi=\pi$ have energies close to the superconducting gap. The subgap bound state then gives rise to a resonance in transmission above the gap~\cite{zhao_phase_2003,zhao_heat_2004} similarly to the transmission across a potential well in standard quantum mechanics. This in turn leads to an increased thermal conductance.

For the case of $p$-wave pairing, the Andreev bound state energies are given by
\begin{equation}
E^\text{ABS}_p(\phi)= \pm|\Delta_0|\sqrt{\tau}\cos\frac{\phi}{2}.
 \label{eq:ABSB}
\end{equation}
From this expression we can infer that $\tau$ only modulates the energy of the Andreev bound state at $\phi\neq\pi$ and that there is always a pair zero-energy bound states at $\phi=\pi$. These robust zero-energy states are topologically protected Majorana bound states, i.e. they are their own particle-hole conjugate. It is the presence of these Majorana bound states that prevents the resonance physics discussed for the $s$-wave case around $\phi=\pi$ and, thus, is directly responsible for the robust minimum of the thermal conductance at phase difference $\phi=\pi$ (see Fig.~\ref{fig:1dbhasDOS}).

An alternative explanation why the transport signatures of $s$-wave and $p$-wave junctions differ from each other significantly only in the tunneling limit can be obtained by analyzing the scattering processes at energies above the superconducting gap. In the case of a transparent junction, $\tau=1$, particles can be backscattered only via Andreev-like reflections where an incoming electronlike quasiparticle with momentum $+k_F$ is converted into a holelike quasiparticle with momentum $+k_F$ and vice versa. For a finite barrier, additional normal backscattering processes become possible where an electronlike quasiparticle at $+k_F$ is scattered to $-k_F$ without charge conversion. Due to the linear momentum dependence of the $p$-wave order parameter, the backscattered mode is thus effectively phase shifted. This is not the case for the momentum-independent $s$-wave pairing potential. The presence or absence of this phase shift affects the transmission and results in a different thermal conductance.

So far, we have considered transport in one-dimensional junctions. In the case of two-dimensional junctions, we have to take into account modes with a different angle of incidence $\alpha$. While the mode with $\alpha=0$ displays exactly the physics of the one-dimensional case discussed above, modes impinging under a finite angle of incidence do not give rise to a Majorana bound state as topological protection is lost. The associated Andreev bound states thus approach the superconducting gap as the transmission of the junction becomes small. In evaluating the thermal conductance, an average of all angles of incidence has to be performed. In consequence, the clear distinction between the phase-dependence of $s$-wave and $p$-wave junctions is lost and a maximum of the thermal conductance at $\phi=\pi$ can arise in the $p$-wave case.

\subsection{\label{sec:chisc}Unitary Chiral Pairing}
In this section we are going to discuss a Josephson junction made of unitary chiral superconductors. Possible systems which can realize such an order parameter are, e.g., the superfluid He$^3$ A-phase \cite{vollhardt_superfluid_2013} and the superconductor Sr$_2$RuO$_4$ \cite{mackenzie_superconductivity_2003}. 
In a chiral superconductor, the order parameter is nodal which implies that there is a preferred direction in momentum space. Furthermore, as the order parameter is anisotropic, there is also a preferred axis in spin space. The relative orientation of these two axes is governed by spontaneous symmetry breaking and residual interactions such as spin-orbit coupling. The relative orientation has an impact on the transport properties of our junction. In the following, we are choosing our coordinate system in momentum space such that the $z$-axis points along the nodal direction. Choosing a $d$-vector along the $z$ axis then gives rise to $S_z=0$ pairing while a $d$-vector oriented in the $x-y$ plane leads to equal spin pairing. The explicit form of the $d$-vector is given by~\cite{anderson_generalized_1961}
\begin{align}
\vec d(\vec k)=\frac{\Delta_0}{k_F}\left(k_x +ik_y\right)\vec e_i,
\label{eq:dchiral}
\end{align}
where $i=x,y,z$.
From the above expression we see that even if $\Delta_0$ is gauged real, the $d$-vector has complex components and as a result time-reversal symmetry is broken for an isolated bulk superconductor,
\begin{equation}
TH_{\text{BdG}}(\vec k)T^{-1}\neq H_{\text{BdG}}(-\vec k).
\end{equation}
Due to the symmetry reduction compared to the helical case, the chiral topological superconductors therefore belong to class D of the Cartan classification~\cite{ryu_topological_2010}.
We note that the order parameters associated with Eq.~\eqref{eq:dchiral} do not depend explicitly on $k_z$. The size of the superconducting gap is given by 
\begin{align}
\frac{1}{2}\text{Tr}\left[\hat{\Delta}(\vec k)\hat{\Delta}^\dagger(\vec k)\right]=|\Delta_0|^2\sin^2\gamma,
\label{eq:anis}
\end{align} 
where $\gamma=\arccos(k_z/k_F)$ and, hence, vanishes along the $z$ axis.

\begin{figure}
	\includegraphics[width=0.95\columnwidth]{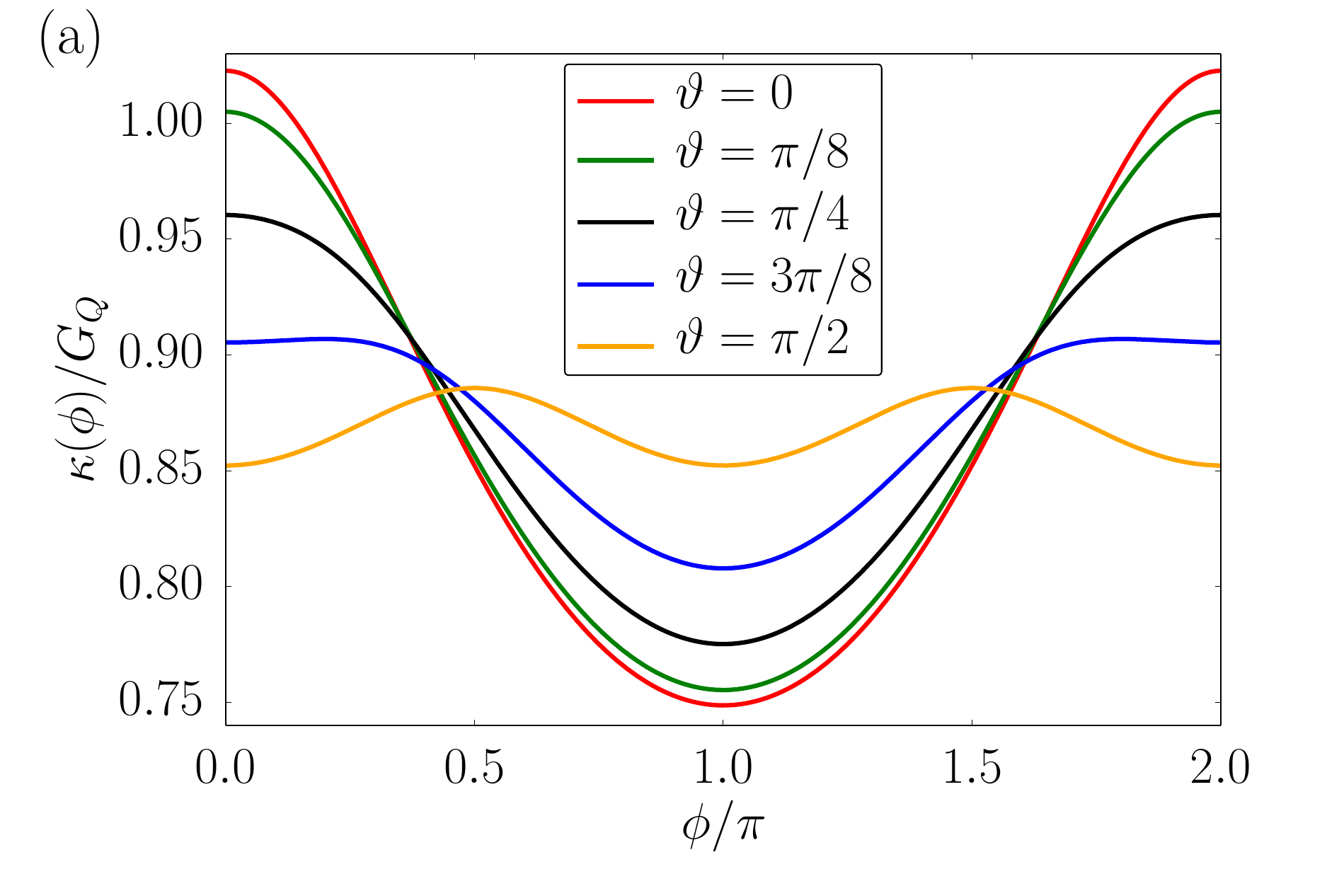}
	\includegraphics[width=0.85\columnwidth]{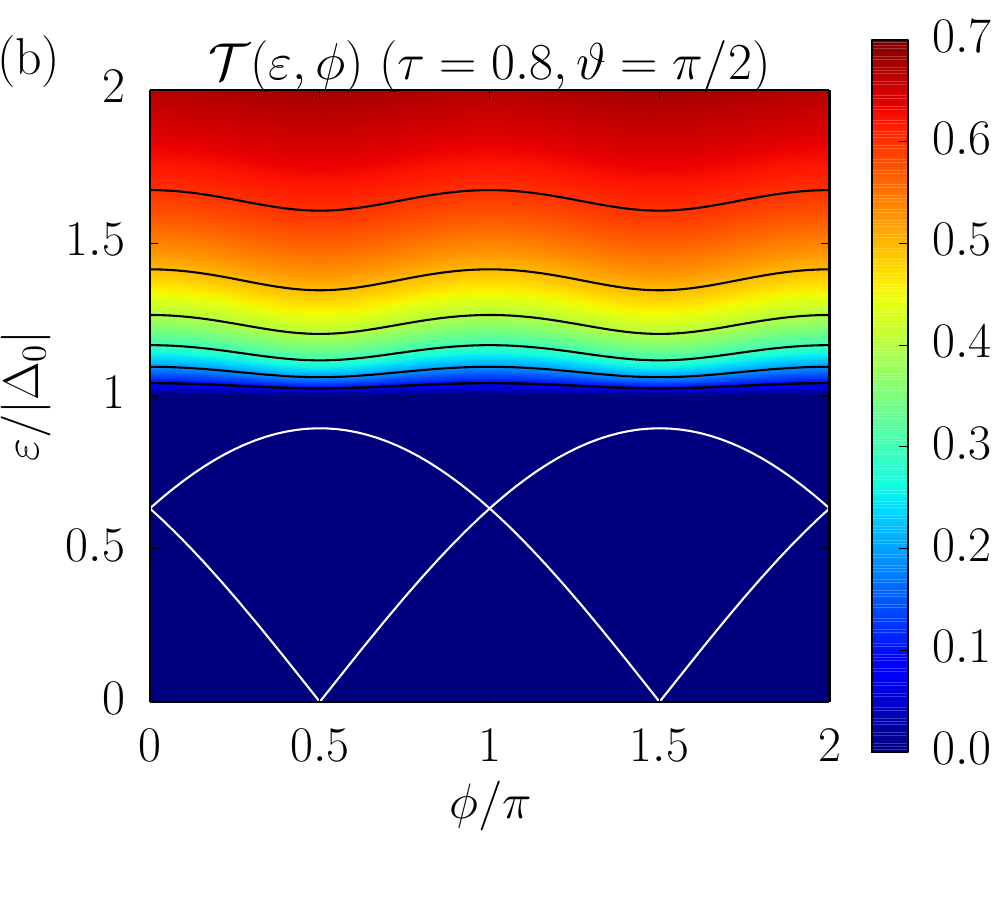}
	\caption{\label{fig:1dahasDOS}(a) Phase-dependent thermal conductance for a unitary chiral one-dimensional SNS junction for different angles $\vartheta$ between the left and right $d$-vector and transmission $\tau=0.8$. (b) Transmission function $\mathcal{T}(\varepsilon,\phi)$ for $\vartheta=\pi/2$ and $\tau=0.8$.
	Parameters are chosen as in Fig.~\ref{fig:1dbhas}.}
\end{figure}
We start our analysis of thermal transport again for the case of a one-dimensional junction. 
Since the order parameter vanishes along the $z$-axis, we consider transport in the $x$ direction without loss of generality. In contrast to the helical system, for the chiral setup the relative orientation of the $d$-vectors affects the transport properties. As shown in Fig.~\ref{fig:1dahasDOS}(a), for a parallel orientation of the two $d$-vectors, the thermal conductance of the chiral system is identical to that of the helical system, i.e., it is $2\pi$-periodic and exhibits a minimum at phase difference $\phi=\pi$. However, if the two $d$-vectors are orthogonal to each other, the periodicity of $\kappa$ is changed from $2\pi$-periodic to $\pi$-periodic with maxima at $\phi=\pi/2$. When the $d$-vectors enclose an intermediate angle $\vartheta$, there is a smooth transition between these two limiting cases.

Physically, this change of periodicity can be traced back to the nontrivial spin structures of Cooper pairs. Let us consider the situation where the $d$-vector of the left superconductor points along the $z$-axis while the $d$-vector of the right superconductor points along the $x$-axis. In this case, Cooper pairs on the left are in an $S_z=0$ state and correspondingly have a spin structure of the form $\up\down+\down\up$ while Cooper pairs on the right are in an $S_x=0$ state with a corresponding spin structure of the form $\up\up-\down\down$. Hence, spin conservation requires the coherent transfer of pairs of Cooper pairs resulting in a reduced periodicity of the thermal conductance.

The change of periodicity is also reflected in the Andreev bound state spectrum shown in Fig.~\ref{fig:1dahasDOS}(b). When the $d$-vector are oriented perpendicularly, the Andreev bound state energies are given by
\begin{align}
	E^\text{ABS}_{1,2}=\pm|\Delta_0|\sqrt{\tau}\cos\left(\frac{\phi}{2}+\frac{\pi}{4}\right),\\
	E^\text{ABS}_{3,4}=\pm|\Delta_0|\sqrt{\tau}\cos\left(\frac{\phi}{2}-\frac{\pi}{4}\right),
\end{align}
giving rise to a $\pi$-periodic spectrum. Furthermore, we find that for $\phi=\pi/2$, one of the Andreev bound states is close to the superconducting gap. As explained above, this leads to a resonance in the transmission above the gap and, thus, to a maximum of the thermal conductance.

We now turn to the case of two-dimensional junctions. Due to the anisotropic order parameter of chiral superconductors, the orientation of the junction with respect to the nodal points of the order parameter plays an important role for the transport properties of the junction. As we have chosen the nodal points to be along the $z$ axis, we can distinguish the two qualitatively different cases of junctions in the $x-y$ and $x-z$ plane.
In the former case, all modes experience the same gap $\Delta_0$ and behave just as in the helical case. As a result, the thermal conductance exhibits a crossover between a minimum and maximum at phase difference $\phi=\pi$ when the transmission is reduced.
For a junction in the $x-z$ plane, it is important to realize that the order parameter~\eqref{eq:dchiral} does not depend on $k_z$ explicitly. Therefore, all modes behave essentially as a copy of the one-dimensional case but with a gap that depends on $k_z$ according to~\eqref{eq:anis}. In consequence, when averaging over all angles of incidence, we find that there is a robust minimum of the thermal conductance at $\phi=\pi$ even for a two-dimensional junctions.

\subsection{\label{sec:favchisc}Nonunitary chiral pairing}
Up to this point we have focussed on helical and chiral order parameters that are unitary. Now, we extend our analysis to the case of a nonunitary chiral order parameter where $\hat\Delta(\vec k)\hat\Delta^\dagger(\vec k)$ is not proportional to the unit matrix any longer. Nonunitary pairing implies that the two spin species exhibit different pairings, up to the extreme case where one species is paired while the other is not. Examples of nonunitary chiral pairing are given by the superfluid A1-phase of He$^3$ which can be stabilized by applying a magnetic field to the A-phase~\cite{vollhardt_superfluid_2013} and by ferromagnetic heavy-fermion superconductors such as UGe$_2$ and URhGe\cite{aoki_review_2019}.
For nonunitary chiral pairing, the superconducting gap becomes maximal in one momentum space direction while it vanishes in the nodal plane perpendicular to this direction. If we choose our coordinate system in momentum space such that the $z$-axis points along the direction where the gap is maximal, the $d$-vector can be written as~\cite{machida_phenomenological_2001}
\begin{equation}
	\vec d(\vec k)=\frac{\Delta_0}{2k_F} \left(k_z,ik_z,0\right)^T.
	\label{eq:A1}
\end{equation}
Just as for the unitary chiral phase time-reversal symmetry is broken,
\begin{equation}
TH_{\text{BdG}}(\vec k)T^{-1}\neq H_{\text{BdG}}(-\vec k),
\end{equation}
and the superconductor falls into the topological symmetry class D~\cite{ryu_topological_2010}. The angular dependence of the superconducting gap is given by
\begin{equation}
\text{Tr}\left[\hat{\Delta}(\vec k)\hat{\Delta}^\dagger(\vec k)\right]= |\Delta_0|^2\left(\frac{k_z}{k_F}\right)^2.
\label{eq:anis2}
\end{equation}

\begin{figure}
	\includegraphics[width=1.0\columnwidth]{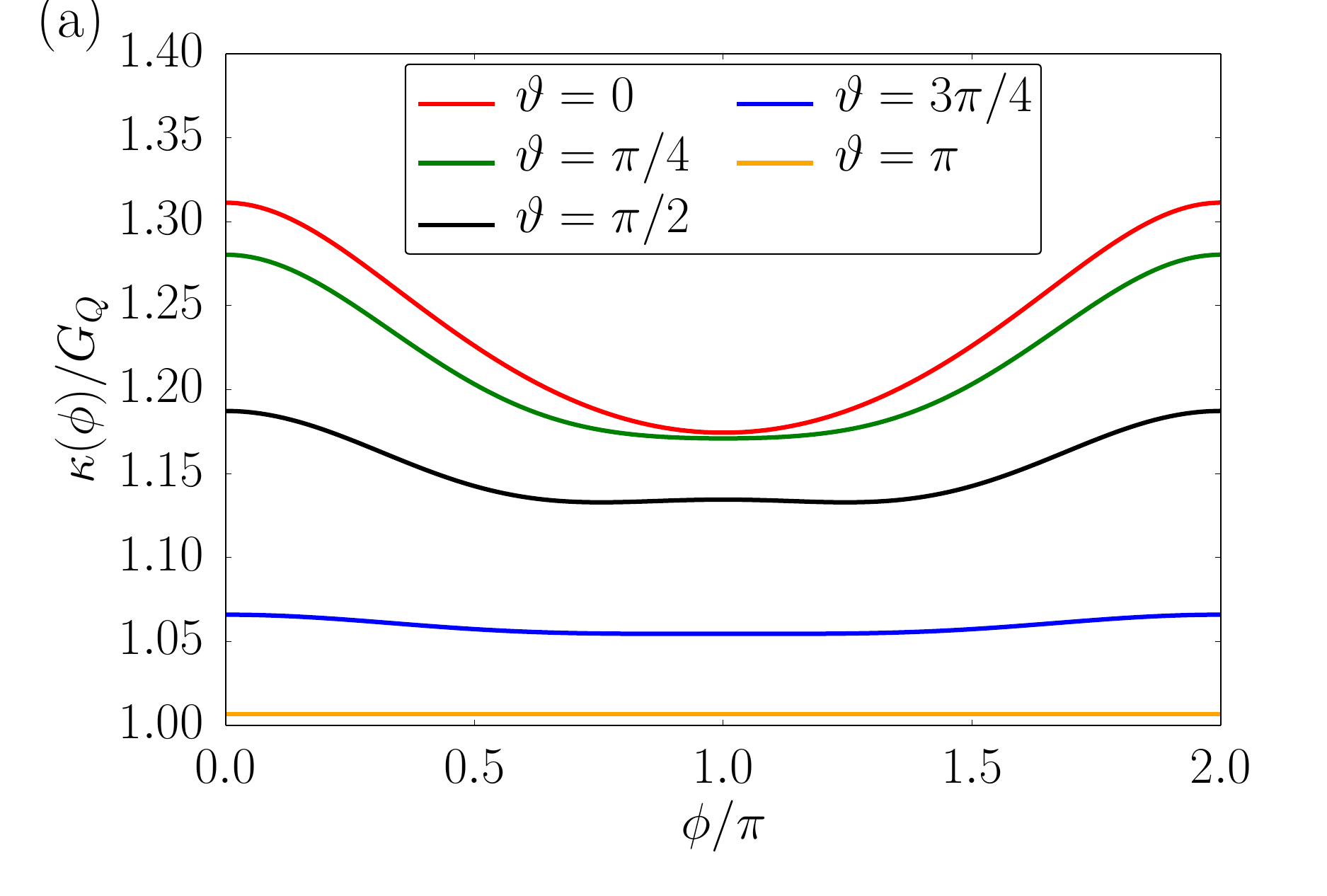}
	\includegraphics[width=1.0\columnwidth]{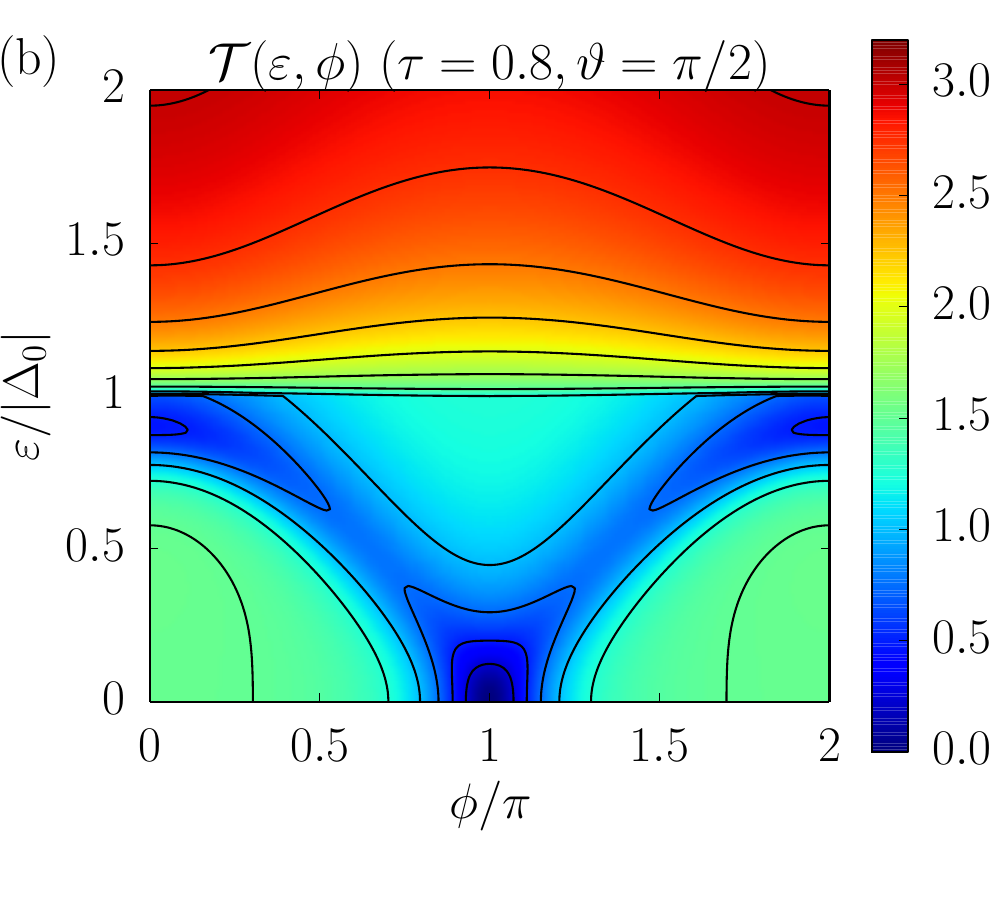}
	\caption{\label{fig:A1} (a) Thermal conductance $\kappa(\phi)$ of a one-dimensional SNS junction with nonunitary chiral pairing for different angles $\vartheta$ between the $d$-vectors on the left and right side for a transmission $\tau=0.8$. (b) Transmission function $\mathcal{T}(\varepsilon,\phi)$ for $\vartheta=\pi/2$ and $\tau=0.8$. We remark that $\mathcal{T}(\varepsilon,\phi)$ accounts for transmission of electron- and holelike quasiparticles of both spins and, hence, is bounded by $4$. Parameters are chosen as in Fig.~\ref{fig:1dbhas}.}
\end{figure}

We are going to analyze thermal transport in a one-dimensional junction of nonunitary chiral superconductors first. If the $d$-vectors on both sides of the junction point in the same direction, the paired spin component contributes to the phase-dependent thermal conductance just as in the case of helical and unitary chiral pairing. At the same time, the unpaired component gives rise to an additional phase-independent thermal conductance $\tau G_Q$ where $G_Q=\pi^2\kB^2 T/(3h)$ denotes the thermal conductance quantum, see Fig.~\ref{fig:A1}(a). If the $d$-vectors point in opposite directions, the thermal conductance becomes phase-independent because the paired component of the left superconductor is unpaired in the right superconductor and vice versa. In consequence, both spin channels just contribute the phase-independent thermal conductance of an NS junction. Finally, for noncollinear orientations of the $d$-vectors, the thermal conductance exhibits significant contributions from higher-order harmonics, i.e. coherent higher-order Cooper pair transfer. This is due to the fact that the paired component in the left superconductor now couples to both, the paired and unpaired component in the right superconductor. As a result, quasiparticles can leak in and out of the Andreev bound states and give rise to multi-Cooper pair transfers. In the  transmission function, this behavior leads to regions of reduced transmission along the broadened position of the Andreev bound states, cf. Fig.~\ref{fig:A1}(b).

We now turn to the case of two-dimensional junctions and focus on the situation with equal $d$-vectors in both superconductors. For a junction oriented in either the $x-z$ or $y-z$ plane all modes experience the same order parameter up to a trivial modulation of its absolute values as described by Eq.~\eqref{eq:anis2}. Subsequently, the thermal conductance behaves qualitatively similar to the one-dimensional case and exhibits a minimum at phase difference $\phi=\pi$ for any value of the transmission $\tau$. This robust behavior can also be understood by considering the energies of the Andreev bound states forming inside the junction.
\begin{equation}
E^\text{ABS}= \pm |\Delta_0|\frac{\left(\frac{k_z}{k_F}\right)^2}{\sqrt{\tau^{-1}-1+\left(\frac{k_z}{k_F}\right)^2}}\cos\frac{\phi}{2}.
\label{eq:A1ABS}
\end{equation}
While the energy in general depends on both $k_z$ and $\tau$, there is a topologically protected zero-energy state at $\phi=\pi$ which, as discussed before, leads to a minimum of the thermal conductance.

\section{\label{sec:con}Conclusion}
We have studied phase-coherent heat transport in short SNS Josephson junctions based on unconventional superconductors and superfluids. In particular, we have analyzed junctions with different spin-triplet $p$-wave pairings namely (i) helical pairing, (ii) unitary chiral pairing and (iii) nonunitary chiral pairing. 
In the case of one-dimensional junctions we found that for helical pairing the thermal conductance is $2\pi$-periodic and always has a minimum at phase difference $\phi=\pi$ connected to the presence of a topologically protected zero-energy Andreev bound state in the junction. For unitary chiral pairing, the periodicity of the thermal conductance can be changed from $2\pi$ to $\pi$ by controlling the relative orientation of the $d$-vectors. Finally, for nonunitary chiral pairing we found a behaviour similar to the helical case but with a phase-independent offset which is connected to the fact that only one spin species is paired.
For two-dimensional junctions we found that they allow for an additional distinction between the different pairings as the interplay between the anisotropy of the order parameter and the junction orientation has a strong impact on the transport properties.

In summary, we have identified phase-coherent heat transport as a valuable tool to probe unconventional superconductivity. Possible directions for future reasearch include the investigation of even more exotic types of pairing that occur, e.g., in superfluid He$^3$ on confined geometries~\cite{levitin_evidence_2019}. Furthermore, it is of interest to investigate how a self-consistent calculation of the order parameter in the junction which is beyond the scope of our present work affects thermal transport properties.

\begin{acknowledgments}
We acknowledge financial support from the Ministry of Innovation NRW via the ``Programm zur Förderung der Rückkehr des hochqualifizierten Forschungsnachwuchses aus dem Ausland''.
\end{acknowledgments}


%

\end{document}